\newif\ifsingle
\twocolumn

\newif\ifFullVersion
\FullVersiontrue 

\ifsingle
\documentclass[12pt,draftclsnofoot, onecolumn]{IEEEtran}		
\else		
\documentclass[10pt,final, twocolumn]{IEEEtran}
\fi

\usepackage{amsmath,amsfonts}
\usepackage{algorithmic}
\usepackage{algorithm}
\usepackage{array}
\usepackage{textcomp}
\usepackage{stfloats}
\usepackage{url}
\usepackage{graphicx}
\usepackage{cite}
\usepackage{times}
\usepackage{amssymb,amsthm}
\usepackage{epsfig,verbatim}
\usepackage{setspace}
\usepackage{color}
\usepackage{epstopdf}
\usepackage{accents}
\usepackage{acronym}
\usepackage[bookmarks,colorlinks]{hyperref}
\usepackage{booktabs}
\usepackage{mathtools}
\usepackage{subfigure}
\usepackage{enumitem}
\usepackage{bbm}

\begin{document}
\acrodef{ula}[ULA]{uniform linear array}
\acrodef{lfm}[LFM]{linear frequency modulation}
\acrodef{2d}[2D]{two-dimension}
\acrodef{siso}[SISO]{single-input single-output}
\acrodef{rcs}[RCS]{radar cross section}
\acrodef{doi}[DOI]{domain of interested}
\acrodef{fmcw}[FMCW]{frequency-modulated continuous-wave}
\acrodef{fista}[FISTA]{fast iterative shrinkage-thresholding algorithm}
\acrodef{resnn}[ResNet]{residual neural network}
\acrodef{awgn}[AWGN]{additional white Gaussian noise}
\acrodef{relu}[ReLU]{Rectified linear unit}
\acrodef{snr}[SNR]{signal-to-noise-ratio}
\acrodef{mse}[MSE]{mean square error}
\acrodef{ssim}[SSIM]{structural similarity index measure}
\acrodef{cs}[CS]{compressed sensing}

\title{Physics-assisted Deep Learning for FMCW Radar Quantitative Imaging of Two-dimension Target}
\author{  
	\IEEEauthorblockN{Zhuoyang Liu$^*$,~\IEEEmembership{Graduate Student Member,~IEEE}, Huilin Xu$^*$,~\IEEEmembership{Graduate Student Member,~IEEE},\\
 Feng Xu$^*$,~\IEEEmembership{Senior Member,~IEEE}\\
	} 
 \IEEEauthorblockA{{$^*$}
 \textit{Key Lab for Information Science of Electromagnetic Wave (MoE)}, 
\textit{Fudan University},
Shanghai 200433, China,\\
}
\IEEEauthorblockA{Email:\{liuzy20,fengxu\}@fudan.edu.cn,~hlxu21@m.fudan.edu.cn}
}

	
	



\maketitle

\begin{abstract}
Radar imaging is crucial in remote sensing and has many applications in detection and autonomous driving. However, the received radar signal for imaging is enormous and redundant, which degrades the speed of real-time radar quantitative imaging and leads to obstacles in the downlink applications. In this paper, we propose a physics-assisted deep learning method for radar quantitative imaging with the advantage of \ac{cs}. Specifically, the signal model for \ac{fmcw} radar imaging which only uses four antennas and parts of frequency components is formulated in terms of matrices multiplication. The learned fast iterative shrinkage-thresholding algorithm with residual neural network (L-FISTA-ResNet) is proposed for solving the quantitative imaging problem. The L-FISTA is developed to ensure the basic solution and ResNet is attached to enhance the image quality. Simulation results show that our proposed method has higher reconstruction accuracy than the traditional optimization method and pure neural networks. The effectiveness and generalization performance of the proposed strategy is verified in unseen target imaging, denoising, and frequency migration tasks.
\end{abstract}
\begin{IEEEkeywords}
Radar quantitative imaging, \ac{cs}, physics-assisted deep learning, L-FISTA-ResNet.
\end{IEEEkeywords}

\section{Introduction}


Radar imaging has attracted considerable attention in remote sensing, detection, and autonomous driving areas \cite{sun2020mimo}, due to its capability of penetrating clothes and clouds while having strong reflections for metallic materials. Quantitative radar imaging plays a key role in accurately measuring and characterizing targets and enhances the understanding of the physical properties of objects and environments. 
However, in the case of high-resolution imaging, we need a high sampling rate which will lead to huge memory consumption and be hard to deal with \cite{yang2013compressed}. 

Compressed sensing has emerged as a promising approach for radar imaging, enabling the reconstruction of the target posture and physical properties of materials \cite{potter2010sparsity}.
A large body of research employs \ac{cs}-based techniques with sparsity to recover high-resolution target images from limited radar data, which offer stability and good interpretability \cite{ccetin2014sparsity,potter2010sparsity,xu2022sparse,yang2013compressed}.
Unlike the traditional imaging methods, the \ac{cs}-based approach, such as \ac{fista}\cite{beck2009fast}, first models the imaging process as an optimization problem and then utilizes the regularization technique to constrain the sparsity and reconstruct the image of the interested target \cite{li2022efficient,liu2021image}.
However, the iterative nature of the \ac{fista} \cite{beck2009fast,florea2020generalized} often leads to poor computational efficiency, particularly in scenarios involving large-scale and high-resolution imaging tasks. These iteration-based algorithms exhibit high sensitivity to hyperparameter selection and require significant effort for tuning. Moreover, they may encounter challenges in capturing nonlinear relationships behind complex radar data.

In recent years, deep learning approaches have made remarkable achievements in various fields \cite{he2016deep,zhao2019object,zhu2021deep}. Naturally, neural networks, characterized by their powerful feature representation and parallel processing capabilities, have emerged as novel tools for radar imaging \cite{zhang2021synthetic,gao2018enhanced,cheng2020compressive}. Generally, neural networks are employed to establish a mapping between raw data and imaging results, facilitating faster speed and more accurate physical properties reconstruction. Nevertheless, several challenges remain. Firstly, these models rely heavily on enormous training samples, which are often scarce in real-world scenarios. Moreover, as black-box models, they lack good interpretability. Therefore, several efforts have been made to integrate physics-assisted optimization algorithms with neural networks in order to balance imaging quality, computational efficiency, and model interpretability. Wang \textit{et al.} propose LFISTA-Net, which incorporates \ac{fista} \cite{beck2009fast} and a deep neural network for precise reconstruction in mmW 3-D holography, and achieves high speed and low computational cost \cite{wang2021lightweight}.  
Xiang \textit{et al.} present FISTA-Net's outstanding performance for diverse imaging tasks, such as  Electromagnetic Tomography (EMT) with strong generalization and noise robustness \cite{xiang2021fista}.

Inspired by previous works, we consider the quantitative imaging for the \ac{2d} target with a sparse sampling of the raw data. 
In particular, we seek to reconstruct the \ac{rcs} map of the target in case of dealing with sparse sampled signals.
In the following sections, we first introduce the general \ac{fmcw} radar signal model and formulate the optimization problem for \ac{rcs} reconstruction. 
To tackle the corresponding problem, we develop a physics-assisted deep learning method, that combines the benefits of \ac{fista} and the deblurred mechanism of the \ac{resnn}. Specifically, We describe the standard \ac{fista}, which is extended into a learnable architecture, and present the overall architecture of the proposed L-\ac{fista}-\ac{resnn}. Finally, we provide extensive comparison experiments to evaluate the performance of our method on the synthesized dataset. 

\section{System model and problem formulation}
\label{proble_form}

\subsection{General FMCW Signal Model}
\begin{figure}[htbp]
     \centering
     \includegraphics[width=0.95\linewidth]{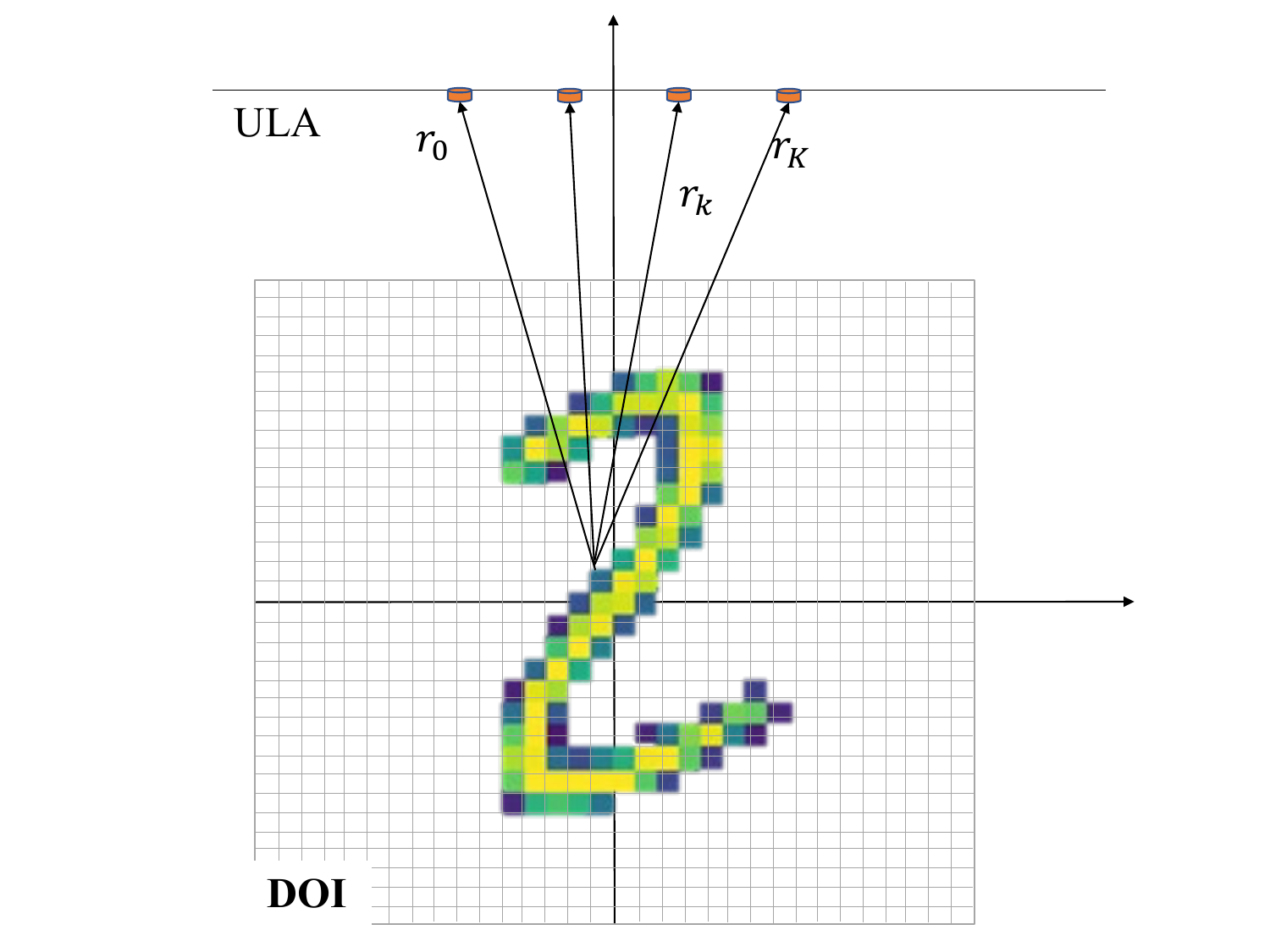}
     \caption{FMCW radar quantitative imaging configuration.}
     \label{fig:scene}
 \end{figure}

We consider the \ac{fmcw} radar quantitative imaging for the \ac{2d} scenario, as shown in Fig.~\ref{fig:scene}. The \ac{2d} target is put in the center of the \ac{doi}, and the \ac{ula} consisting of $K$ antennas is deployed at $2~m$ from the coordinate center, parallel to the x-axis. Specifically, we mesh the \ac{2d} target into $M$ points and all antennas are \ac{siso} and work in \ac{lfm} mode. 
Thus, the received radar beat signal of the $k=1,...,K$ antenna at time step $t$ is given by
\begin{equation}
\label{radar_echo_t}
    \begin{aligned}
    \centering
    s(t,k)=\sum_{m=1}^{M}\epsilon_m\exp\left(-j2\pi (f_0+K_r t) \cdot 2\tau_{k,m}\right),~ 0\leq t \leq T,
    \end{aligned}
\end{equation}
where $T$ is the chirp duration, $\epsilon_m$ is the \ac{rcs} of the $m$-th point of the \ac{2d} target, $\tau_{k,m}$ is the time delay of the echo from the $k$-th antenna to the $m$-th point, $f_0$ is the starting frequency, and $K_r$ is the rate of the frequency sweep of each chirp. Let $B$ denote the bandwidth and $N_f$ be the number of frequency grids, then (\ref{radar_echo_t}) can be further transformed into the frequency domain, expressed as
\begin{equation}
\label{radar_echo_f}
    \begin{aligned}
        \centering
        s(n,k)= \sum_{m=1}^{M}\epsilon_m\exp\left(-j2\pi \left(f_0+\frac{B}{N_f}n\right) \cdot2\frac{R_{k,m}}{c}\right)&,\\
        n=0,...,N_f-1&,
    \end{aligned}
\end{equation}
 where $c$ is the speed of light, and time delay $\tau_{k,m}$ is represented as $\frac{R_{k,m}}{c}$ with $R_{k,m}$ being the distance between the $m$-th point and the $k$-th antenna. For analysis, we mesh the \ac{doi} into $P$ grids and each grid only contains one point of the \ac{2d} target. 
Therefore, the radar echo of each antenna is rewritten as
 \begin{equation}
\label{radar_echo_f_mesh}
    \begin{aligned}
        \centering
        s(n,k)= \sum_{p=1}^{P}\epsilon_p\exp\left(-j2\pi \left(f_0+\frac{B}{N_f}n\right) \cdot2\frac{R_{k,p}}{c}\right)&,\\
        n=0,...,N_f-1&,
    \end{aligned}
\end{equation}
where $R_{k,p}$ is the distance between the $p$-th grid and the $k$-th antenna, and $\epsilon_p$ is the \ac{rcs} of the $p$-th grid. Obviously, if the $p$-th grid contains the $m$-th point of the \ac{2d} target, it satisfies $\epsilon_p = \epsilon_m$, otherwise $\epsilon_p = 0$.
 Next, we assemble the \ac{rcs} corresponding to each grid of the \ac{doi} into a vector $\boldsymbol{\epsilon}=[\epsilon_1,...,\epsilon_P]^T\in\mathbbm{R}^{P\times 1}$, and collect the distance between each grid to the $k$ antenna into $\boldsymbol{r}_k=[R_{k,1},...,R_{k,P}]^T\in\mathbbm{R}^{P\times 1}$. 
 For the $k$-th antenna, the sensing matrix $\boldsymbol{A}_k$ is denoted by
 \begin{equation}
 \label{sensing_matrix_k}
     \begin{aligned}
         \centering
         \boldsymbol{A}_k = \exp\left(-j4\pi \frac{\boldsymbol{f}\boldsymbol{r}_k^H}{c}\right)
     \end{aligned}
 \end{equation}
 where $\boldsymbol{f}=[f_1,..., f_{N_f}]^T\in\mathbbm{R}^{N_f\times 1}$ is the frequency sweep vector with $f_n=f_0+\frac{B}{N_f}(n-1)$. Then, by concatenating the sensing matrix $\boldsymbol{A}_k\in\mathbbm{C}^{N_f\times P}$ of each antenna into a big matrix $\boldsymbol{A}=[\boldsymbol{A}_1^H,...,\boldsymbol{A}_K^H]^H\in\mathbbm{C}^{N_fK\times P}$, the radar echo $\boldsymbol{s}\in\mathbbm{C}^{N_fK\times 1}$ in (\ref{radar_echo_f_mesh}) with \ac{awgn} $\boldsymbol{v}$ is represented by matrix forms, 
 \begin{equation}
 \label{radar_echo_S}
     \begin{aligned}
         \centering
         \boldsymbol{s}=\boldsymbol{A}\boldsymbol{\epsilon}+\boldsymbol{v}.
     \end{aligned}
 \end{equation}
 
The generated radar echo in (\ref{radar_echo_S}) is huge and has redundant information about the target in these $K$ channels. Therefore, by utilizing the sparse sampling technique for antenna channels and wide-band frequency samples, we can reduce the original radar echo in (\ref{radar_echo_S}) into an affordable memory consumption. Particularly, the number of antennas is reduced to $K=4$, and the interval distance of two antennas is $\frac{c}{2f_0}$. The real-time sample rate for one chirp is defined as $f_{ADC}=\frac{B}{N_f}$ with $N_f=50$ to ensure the bandwidth of the down-sampled signal is the same as the original one.

\subsection{Problem Formulation}
In radar quantitative imaging, the goal is to recover the \ac{rcs} map from the received radar signal which is the superposition of the scatterings from the whole \ac{doi}. To address this problem, we utilize optimization approaches to find the feasible solution of (\ref{radar_echo_S}).

Obviously, the sensing matrix $\boldsymbol{A}$ satisfies $N_fK<<P$ after down-sampling, so that $\boldsymbol{\epsilon}$ reconstruction is an underdetermined problem that has multiple feasible solutions. However, the number of points for the \ac{2d} target $M$ is much smaller than $P$ due to the space sparsity of the target in the \ac{doi} \cite{potter2010sparsity}.
That is the underdetermined equation in (\ref{radar_echo_S}) can be solved by using \ac{cs}. To this end, the energy function is the evaluated metric of the $\boldsymbol{\epsilon}$ reconstruction, given by
\begin{equation}
\label{energy function}
    \begin{aligned}
        \centering
        E(\boldsymbol{\epsilon})\triangleq \frac{1}{2}||\boldsymbol{s}-\boldsymbol{A\epsilon}||_2^2+\lambda||\boldsymbol{\epsilon}||_1,
    \end{aligned}
\end{equation}
where $\lambda$ is the penalty part for the balance of the sparsity and fidelity to the measurement. This underdetermined equation in (\ref{energy function}) is always solved by rewriting it as the following optimization problem
\begin{equation}
\label{estimated_rcs}
    \begin{aligned}
        \centering
        \hat{\boldsymbol{\epsilon}} = \min_{\boldsymbol{\epsilon}} E(\boldsymbol{\epsilon}),
    \end{aligned}
\end{equation}
where $\hat{\boldsymbol{\epsilon}}$ is the reconstructed $rcs$ map. The \ac{cs} algorithm in (\ref{estimated_rcs}) for solving (\ref{radar_echo_S}) is to minimize the reconstructed error while imposing the space sparsity of the target. To be concrete, we use the $L_2$ norm to evaluate the quantitative imaging error and the $L_1$ norm for the sparsity constraint. Finally, we have the constrained minimization problem in (\ref{estimated_rcs}) and propose a physics-assisted deep learning method to address it, which will be detailed in the next section.

\begin{figure*}[!ht]
    \centering
    \includegraphics[width=0.85\linewidth]{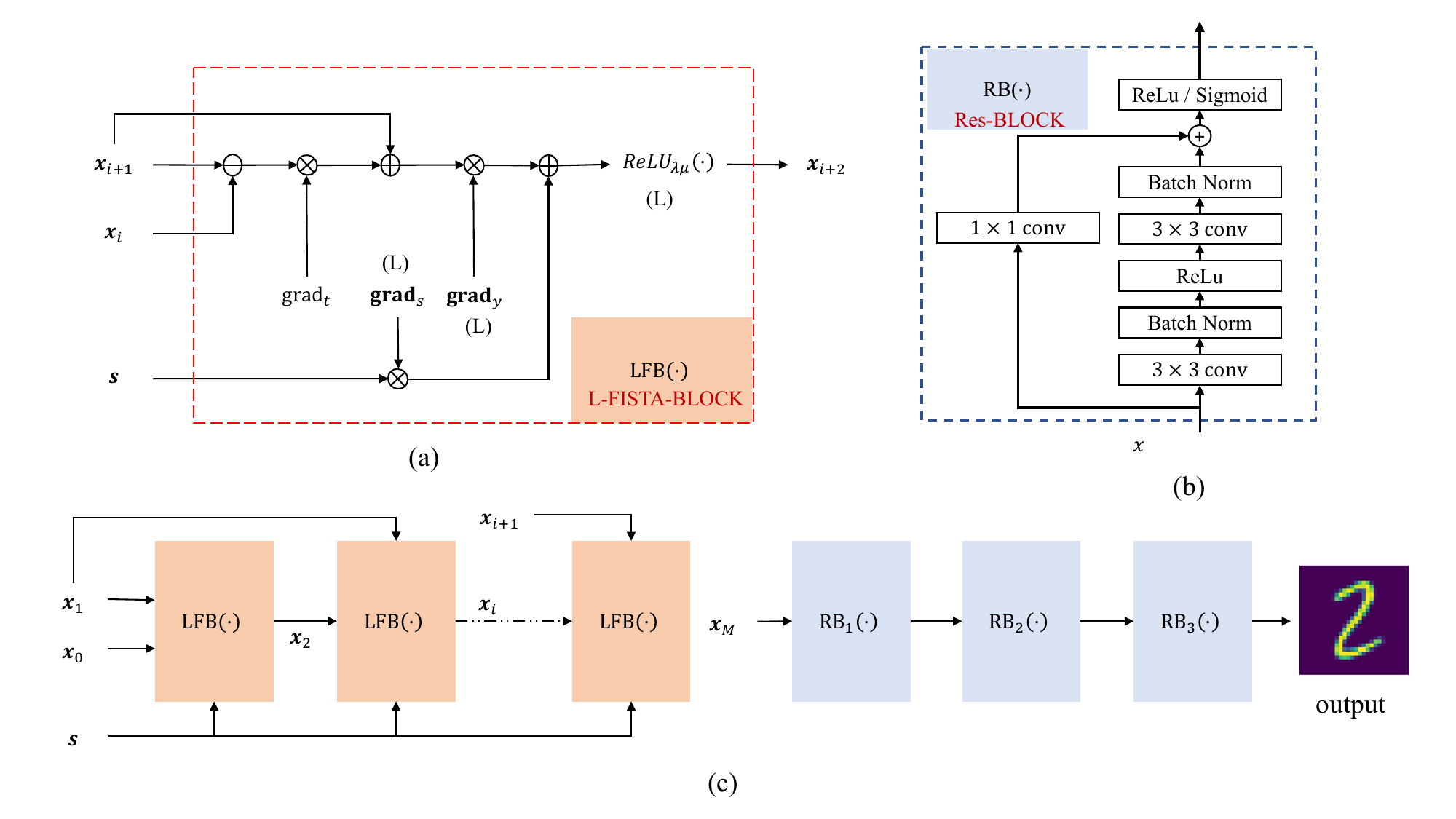}
    \caption{L-\ac{fista}-\ac{resnn} architecture, (a) L-\ac{fista}-BLOCK, (b) Res-BLOCK and (c) pipeline of L-\ac{fista}-\ac{resnn} for quantitative imaging. }
    \label{fig:learn-FISTA-RES}
\end{figure*}

\section{Learn-FISTA-ResNet for quantitative imaging}
\label{l-far}
In this section, we develop the physics-assisted deep learning method for solving the \ac{fmcw} radar quantitative imaging. We begin with the traditional \ac{fista} and then extend it to the learn-\ac{fista}-\ac{resnn}.

\subsection{The Preliminaries of the FISTA}
The quantitative imaging, modeled as a constrained minimization problem in (\ref{estimated_rcs}), is a linear inverse problem. Here we introduce the standard fixed-step \ac{fista} \cite{beck2009fast} to address it, as shown in Algorithm.~\ref{al1_fista}.
\begin{algorithm}[htbp]
    \caption{\ac{fista}}
    \renewcommand{\algorithmicrequire}{\textbf{Input:}}
    \renewcommand{\algorithmicensure}{\textbf{Output:}}
    \label{al1_fista}
    \begin{algorithmic}[1]
        \REQUIRE \leavevmode \\
        Initialize: $t_0=1$, $\mu=1/Lips$, $\boldsymbol{A}$, $\boldsymbol{x}_0=\boldsymbol{0}$,$\boldsymbol{x}_1=\boldsymbol{0}$;\\
        \FOR{$i=0:M-2$}
        \STATE $t_{i+1}=\frac{\sqrt{1+4t_i^2}}{2}$;
        \STATE $\boldsymbol{y}_{i+1}=\boldsymbol{x}_{i+1}+\frac{t_i-1}{t_{i+1}}(\boldsymbol{x}_{i+1}-\boldsymbol{x}_i)$
        \STATE $\boldsymbol{x}_{i+2}=(\boldsymbol{I}-\mu \boldsymbol{A}^H\boldsymbol{A})\boldsymbol{y}_{i+1}+\mu\boldsymbol{A}^H\boldsymbol{s}$
        \STATE $\boldsymbol{x}_{i+2}=\mathcal{S}_{\lambda\mu}(\boldsymbol{x}_{i+2})$
        \STATE $\mathcal{S}_{\lambda\mu}(\boldsymbol{x})=\mathrm{sign}(\boldsymbol{x})\max(|\boldsymbol{x}|-\lambda\mu,0)$
        \ENDFOR
        \ENSURE \leavevmode \\
        The estimated \ac{rcs}: $\boldsymbol{\epsilon}=\boldsymbol{x}_{M}$.
    \end{algorithmic}
\end{algorithm}

To accomplish the basic gradient descent during iterations, we utilize the smallest Lipschitz constant $Lips$ of the gradient $\nabla (\frac{1}{2}||\boldsymbol{s}-\boldsymbol{A\epsilon}||_2^2)$ to represent the basic shrinkage step $\mu$ with \cite{beck2009fast}
\begin{equation}
\label{Lips}
    \begin{aligned}
        \centering
        \frac{1}{\mu}=Lips=\frac{\boldsymbol{A}^H\boldsymbol{A}}{\lambda_{max}},
    \end{aligned}
\end{equation}
where the shrinkage step $\mu$ depends on the sensing matrix $\boldsymbol{A}$. Let $\boldsymbol{x}_i$ for $i = 0,..., M-2$ be defined as the estimated \ac{rcs} map during the iterative loop with $M-1$ being the maximum iteration.
To start \ac{fista}, the inputs $\boldsymbol{x}_0$ and $\boldsymbol{x}_1$ are initialized as zero vectors; and other parameters $t_0$ and $\mathcal{S}_{\lambda\mu}(\cdot)$ in Algorithm.~\ref{al1_fista} are defined as those in \cite{beck2009fast}.

However, it's challenging to find the best penalty parameter $\lambda$ for the specific scenario, and the fixed-step \ac{fista} always requires immense iterations for convergence. These obstacles motivate learnable \ac{fista} and deep learning methods, which will be explained in the next section.

\subsection{L-FISTA-ResNet Architecture}
We convert the standard fixed-step \ac{fista} into a learnable architecture L-\ac{fista} with the following definitions:
\begin{equation}
    \begin{aligned}
        \centering
        \mathrm{grad}_t &= \frac{t_i-1}{t_{i+1}},\\
        \boldsymbol{\mathrm{grad}}_y&=(\boldsymbol{I}-\mu \boldsymbol{A}^H\boldsymbol{A}),\\
        \boldsymbol{\mathrm{grad}}_s&=\mu \boldsymbol{A}^H.
    \end{aligned}
\end{equation}
Based on the expressions of gradients for different temporary functions, the pipeline of the L-\ac{fista} is shown in Fig.~\ref{fig:learn-FISTA-RES}(a). L-FISTA-BLOCK is designed as rewriting each iteration of the \ac{fista} into the learnable form. Specifically, the shrinkage step $\mu$ and penalty parameter $\lambda$ are learned by training. The non-differentiable shrinkage operator $\mathcal{s}_{\lambda\mu}$ is replaced by the \ac{relu} function with the threshold $\lambda\mu$ being learnable. In summary, the threshold of the \ac{relu} function and two gradients $\boldsymbol{\mathrm{grad}}_y$, $\boldsymbol{\mathrm{grad}}_x$ are learned automatically during training rather than by handcrafted design. In the forward process, the cascaded L-\ac{fista} blocks perform multiple iterations and then the output is considered as the convergent solution.

Since we use the down-sampled radar echo as the measured signal, the reconstructed \ac{rcs} map will be blurred and diverged. Combining with the concept of the \ac{resnn} \cite{he2016deep}, the residual blocks can help with deblurred the damaged images. We then refine the initial result obtained from the cascaded L-\ac{fista} blocks with stacked residual blocks to generate the final reconstructed images. Fig.~\ref{fig:learn-FISTA-RES}(b) shows the implementation of the standard Res-BLOCK which is designed as the deblurring procedure.

To sum up, we propose the L-\ac{fista}-\ac{resnn} to achieve the quantitative imaging, where the initial result is generated by L-\ac{fista} and the improved image is obtained by the \ac{resnn}, as shown in Fig.~\ref{fig:learn-FISTA-RES}(c). From the perspective of physics-assisted deep learning techniques, we develop the L-\ac{fista} to ensure the basic solution of the radar quantitative imaging and leverage the \ac{resnn} to enhance the imaging quality for weakening the impact of radar signal down-sampling.

\subsection{Loss Function}

Based on the L-\ac{fista}-\ac{resnn} architecture, we introduce the loss function design in terms of the objective function in \ref{estimated_rcs}. 
We consider the following hybrid loss function:
\begin{equation}
\label{loss_function}
\begin{aligned}
    \centering
    L=\|\boldsymbol{\epsilon}-\hat{\boldsymbol{\epsilon}}\|_2^2+\lambda_1\|\boldsymbol{\epsilon}-\hat{\boldsymbol{\epsilon}}\|_1+\lambda_2\|\boldsymbol{s}-\boldsymbol{A}\hat{\boldsymbol{\epsilon}}\|_2^2,
\end{aligned}
\end{equation}
where the first term uses the $L2$ norm to denote the fidelity to the measurement, and the second term applies the $L1$ norm  between the ground truth and the predicted result to evaluate the sparsity. The last term in (\ref{loss_function}) represents the reconstruction error between the radar echo generated from the prediction results and the measured received signal. To balance the fidelity, sparsity, and reconstruction error, the penalty parameters are chosen empirically $\lambda _1 = 0.1$ and $\lambda _2 = 0.05 $.

\section{Experiments}
In this section, we begin with a brief  introduction of the dataset, metrics, and implementation details. Then quantitative and qualitative experimental results are presented to verify the effectiveness of the proposed method compared with several baselines. Finally, we evaluate the generalization ability of methods on unseen samples of diverse shapes of objects, SNRs, and center frequencies.
\subsection{Experimental Setups}
\textbf{Dataset and Metrics}: We conducted experiments on the widely used MNIST handwritten digit dataset.  Parameters of radar echo generation are listed in Table \ref{PARAMETERS FOR Radar ECHO Generation}.
 We randomly select 2000 samples, with 800 for training, 200 for validation, and 1000 for testing. Two widely used metrics, \ac{mse} and \ac{ssim}, are chosen to evaluate the imaging quality. 

\begin{table}[htbp]
\renewcommand{\arraystretch}{1.3}
\caption{PARAMETERS FOR Radar ECHO Generation.}
\setlength{\tabcolsep}{3mm}
\begin{center}
\begin{tabular}{c|c|c}
\hline
Notation & Parameter &  Quantity\\
\hline
$K$ &  number of  antennas & 4 \\
$N_f$ & number of frequency grids & 50 \\
$p$ & number of  DOI grids & 784\\
$f_0$ & initial frequency &  30 GHz\\
$B$ & bandwidth & 5 GHz\\
\hline
\end{tabular}
\label{PARAMETERS FOR Radar ECHO Generation}
\end{center}
\end{table}

\textbf{Baselines}: Several baselines are chosen to highlight key components of the proposed L-\ac{fista}-\ac{resnn}.
\begin{itemize}
\item \textbf{FISTA}: \ac{fista} \cite{beck2009fast} is a powerful optimization algorithm used for sparse signal recovery and image reconstruction, as shown in Algorithm.~\ref{al1_fista}.
\item \textbf{FISTA-ResNet}: Compared with L-\ac{fista}-\ac{resnn}, we discard the two learnable hyperparameters in the L-\ac{fista} block and instead set them as fixed values.
\item \textbf{DNN}: We replaced the stacked L-\ac{fista} blocks in L-\ac{fista}-\ac{resnn} with simple fully connected layers.
\end{itemize}

\textbf{Implementation details}: We train models with Adam optimizer with a min-batch of 16 for 100 epochs on a single NVIDIA GeForce RTX 3090 GPU. With an initial learning rate of  1e-2, We reduce the learning rate by 0.1 when the validation loss stops decreasing for over 10 epochs. Particularly, For L-\ac{fista}-\ac{resnn}, the number of L-\ac{fista} blocks is set to 20. For \ac{fista}, the maximum number of iterations is 2000. For \ac{fista} and \ac{fista}-\ac{resnn}, hyperparameter $\lambda$  is selected from [0.001, 0.005, 0.01, 0.05, 0.1] and set to the value that achieves the best performance on the training set. Thus $\lambda$ is set to 0.001 and 0.01 respectively for \ac{fista} and \ac{fista}-\ac{resnn}.

\subsection{Result Analysis}
\begin{figure}[htbp]
\centerline{\includegraphics[scale=0.18]{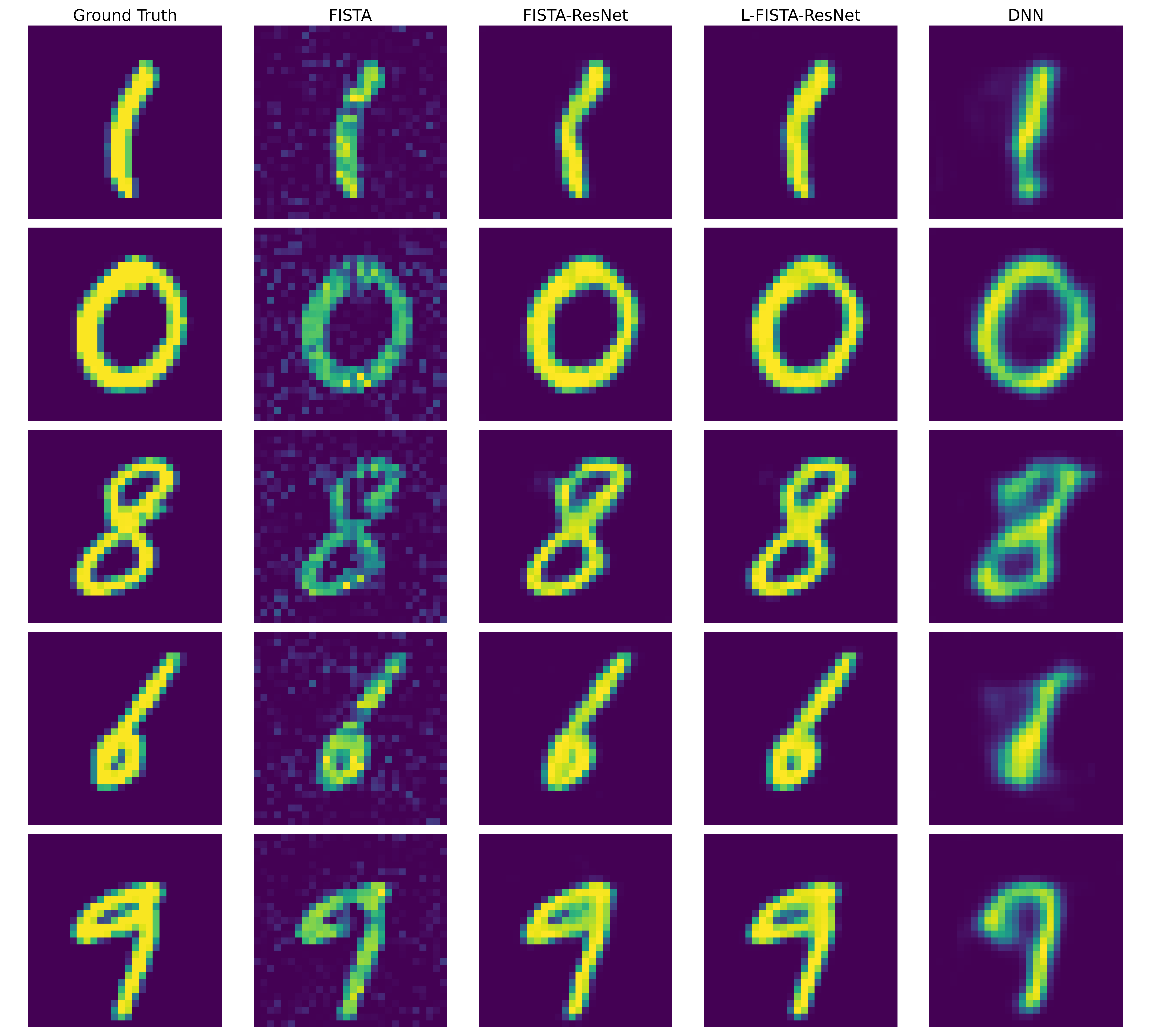}}
\caption{Reconstruction results of MNIST dataset by different methods.}
\label{fig: vis_compared_resutls}
\end{figure}

Performance comparisons of four methods are listed in Table \ref{Performance of four methods on MNIST dataset}. Besides, we also report model parameters and inference time per sample to evaluate the computational efficiency. Our method outperforms the baseline methods by a large margin. On the one hand, due to parallel computation and the powerful fitting ability of neural networks, the proposed method has higher efficiency and better performance compared to the traditional \ac{fista} algorithm. On the other hand, the learnable \ac{fista} block effectively helps maintain great performance while alleviating the dependence on a large number of samples. 
\begin{table}[htbp]
\renewcommand{\arraystretch}{1.3}
\caption{Performance of four methods on MNIST dataset. The bold indicates the best performance.}
\setlength{\tabcolsep}{2.3mm}
\begin{center}
\begin{tabular}{c|c|c|c|c}
\hline
Model & MSE & SSIM & Param. & Time  \\
\hline
\ac{fista} & 0.0124 & 0.872 & $-$ & 7.434 s\\
\ac{fista}-\ac{resnn} & 0.0065 & 0.925 & 7.41 K & 0.060 s \\
L-\ac{fista}-\ac{resnn} & \textbf{0.0049 }& \textbf{0.945} & 7.45 K & 0.051 s \\
DNN & 0.0263 & 0.661 &13.0 K & 0.004 s\\
\hline
\end{tabular}
\label{Performance of four methods on MNIST dataset}
\end{center}
\end{table}
Qualitative results are shown in Fig. \ref{fig: vis_compared_resutls}.

To verify the generalization ability of the proposed model to unseen samples, we conduct a series of experiments. In the later experiments, the models are still trained on the MNIST dataset, but are tested on samples synthesized from targets of different shapes or with different simulation settings.

\begin{figure}[htbp]
\centerline{\includegraphics[scale=0.34]{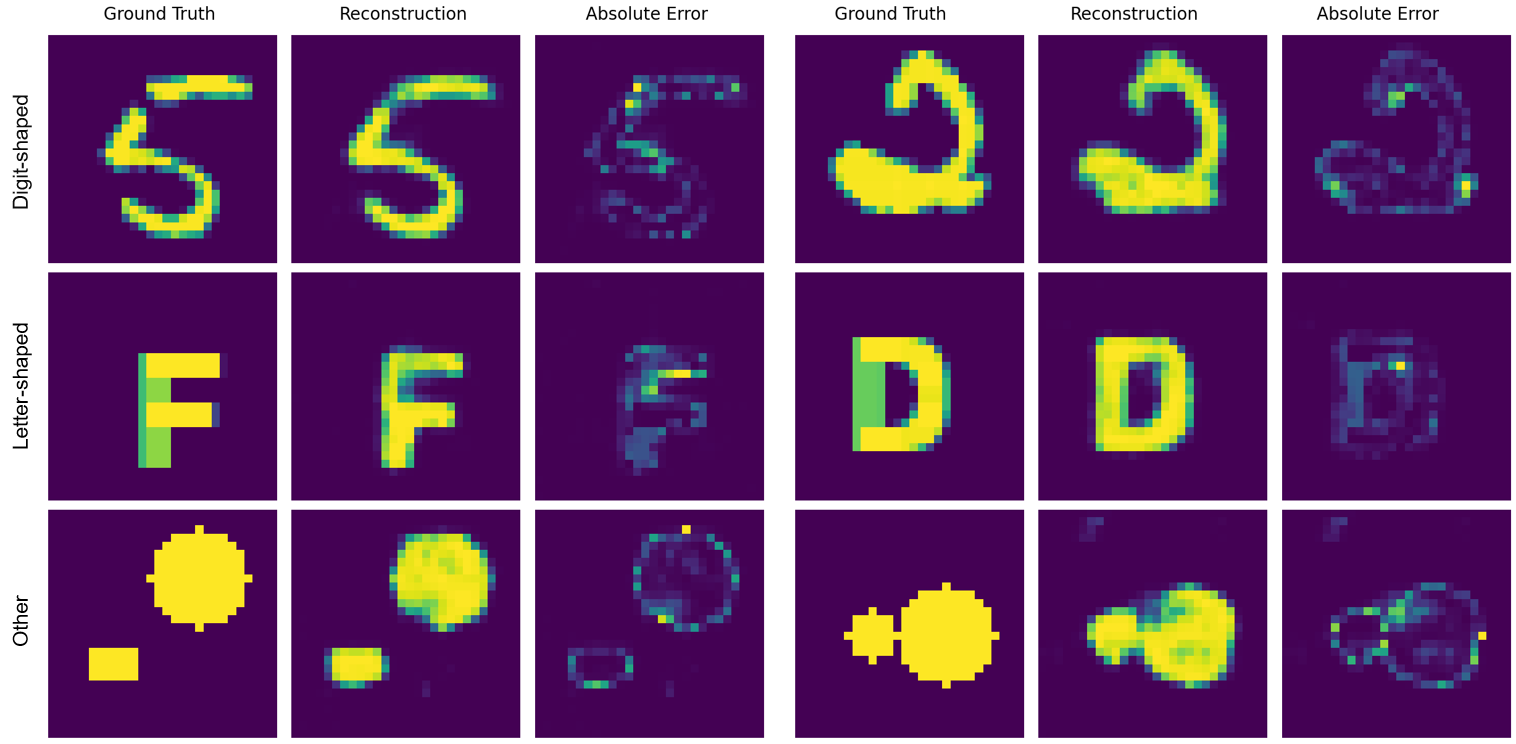}}
\caption{ Reconstruction results of different-shaped objects with L-\ac{fista}-\ac{resnn} trained on MNIST dataset.}
\label{fig:vis_shape}
\end{figure}

We randomly selected several letter-shaped targets and targets consisting of classical shapes as test targets. Fig. \ref{fig:vis_shape} shows the reconstruction results of different methods, the first column is ground truths, the second column is results, and the third column is the absolute error between reconstructions and ground truths. Although L-\ac{fista}-\ac{resnn} is trained only on the digit-shaped MNIST dataset, it still achieved satisfactory reconstruction performance on a variety of targets with different shapes. This phenomenon reflects that L-\ac{fista}-\ac{resnn} effectively models the physical mapping between the input echoes and the output \ac{rcs} images.

\begin{figure}[htbp]
\centerline{\includegraphics[scale=0.3]{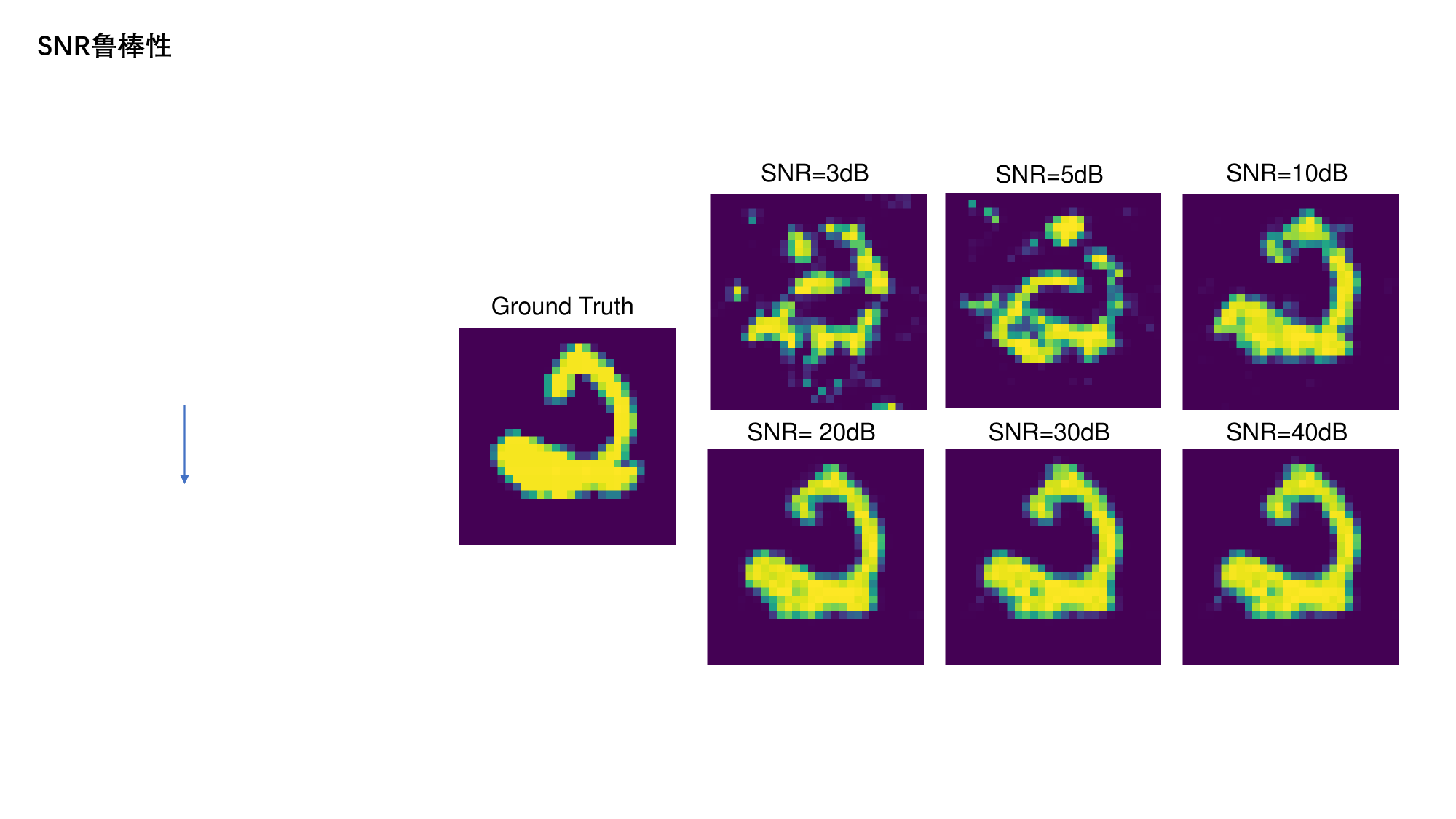}}
\caption{ Reconstruction results by L-\ac{fista}-\ac{resnn} method under different \ac{snr} settings in the test phase.}
\label{fig: vis_SNR}
\end{figure}

\begin{figure}[htbp]
	\centering  
	\vspace{-0.35cm} 
	\subfigtopskip=2pt 
	\subfigbottomskip=2pt 
	\subfigcapskip=-5pt 
	\subfigure[SNR]{
		\label{snr}
		\includegraphics[width=0.48\linewidth]{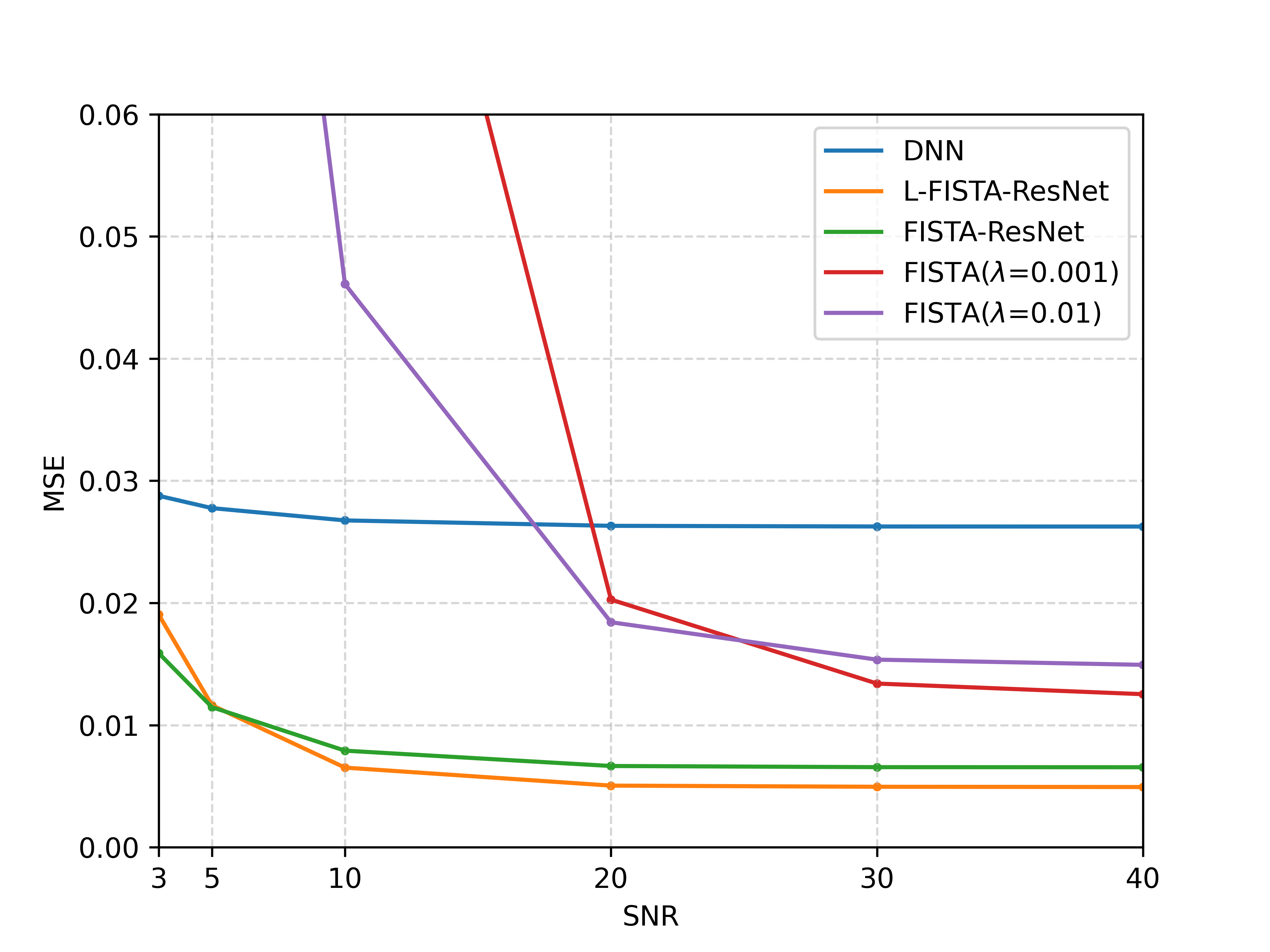}}
  \subfigure[Frequency]{
		\label{freq}
		\includegraphics[width=0.48\linewidth]{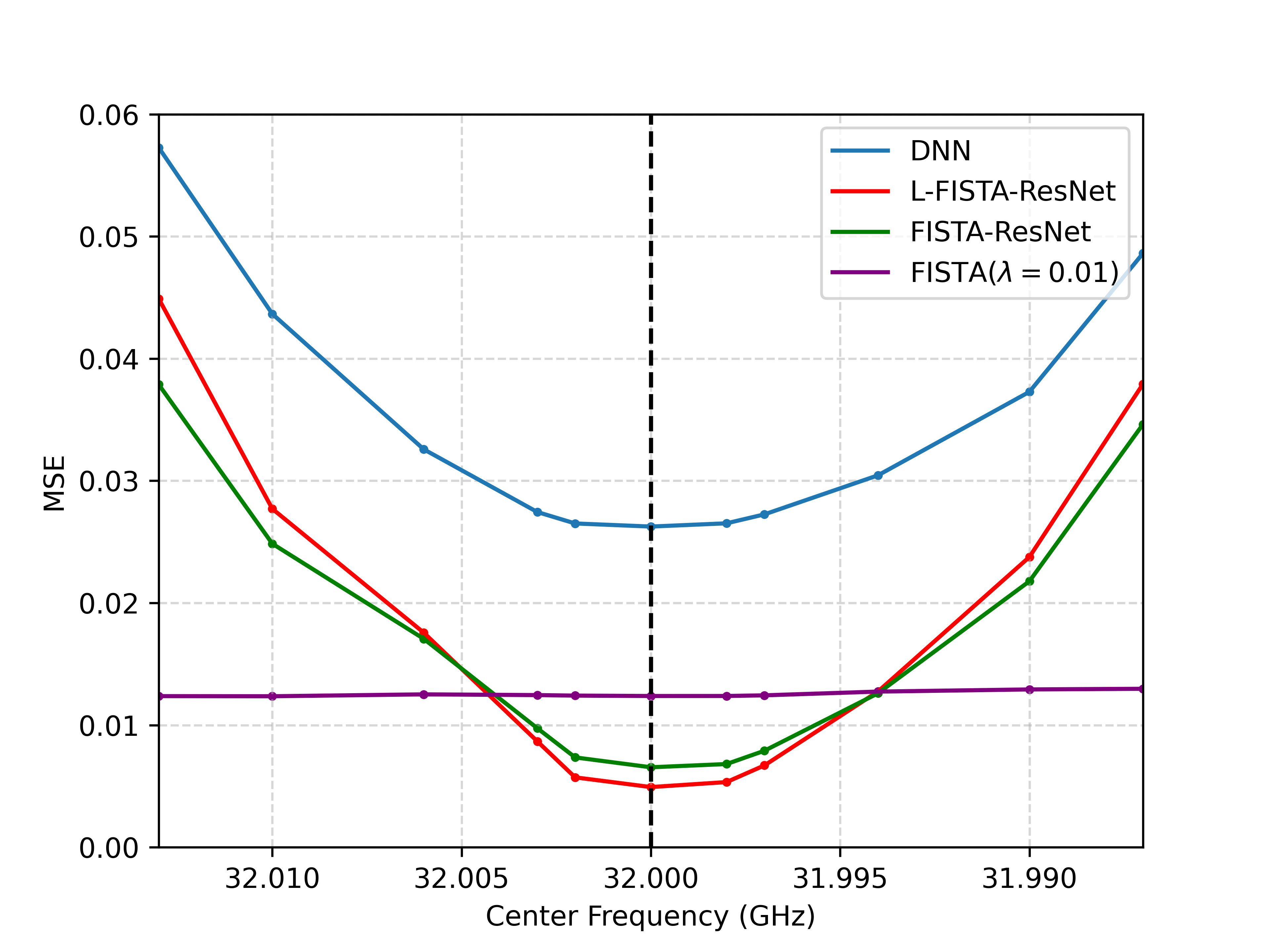}}
	\caption{(a) Performance under different \ac{snr} settings in the test phase.(b) Performance under different center frequency settings in the test phase.}
	\label{fig: SNR_and_Freq}
\end{figure}

For the purpose of assessing the method's robustness to noise, we add additional white Gaussian noise to echoes with varying signal-to-noise ratios to the echo data and reconstruct the \ac{rcs} map using the trained model above in the testing phase. It should be emphasized that the radar echoes used for training are perfect, i.e., noise-free, during the training process. Fig. \ref{fig: vis_SNR} depicts visualizations of reconstructions under different \ac{snr} settings and  Fig. \ref{snr} displays the performance fluctuation when \ac{snr} changes.  The results demonstrate that the proposed L-\ac{fista}-\ac{resnn} has excellent noise robustness and maintains outstanding performance even under challenging low \ac{snr} conditions. Besides, \ac{fista} is sensitive to the hyperparameter $\lambda$. When $\lambda$ decreases, the performance increases at high \ac{snr} but decreases at low \ac{snr}, i.e., the noise robustness worsens

Further, we also tested the trained models on unseen echo samples synthesized at different center frequencies, as shown in Fig. \ref{freq}. It can be seen that our model has great generalization ability even at different degrees of center frequency shifts. \ac{fista}-\ac{resnn} shows slightly better generalization ability because its hyperparameter $\mu$ is adaptively calculated with center frequency changes when $\mu$ is fixed in L-\ac{fista}-\ac{resnn}. Similarly, FISTA has similar performance at different center frequencies. Thus there is a compromise between performance, computational efficiency, and frequency generalization.

\section{Conclusion}

In this work, we achieved the \ac{fmcw} radar quantitative imaging for \ac{2d} targets. With the principle of \ac{cs}, we characterized the quantitative imaging as a constrained minimization problem. To address the constrained problem, we proposed a physics-assisted deep learning approach that combined the advantages of traditional optimization methods and neural networks for \ac{fmcw} radar quantitative imaging. The proposed L-\ac{fista}-\ac{resnn} consists of two key components, the L-\ac{fista} block, and the Residual block. The \ac{fista} component effectively alleviated the dependence of deep learning-based models on a large number of samples while the powerful fitting ability and parallelizable natures of neural networks greatly improved the quality and speed of quantitative imaging. Quantitative and qualitative experimental results showed that L-\ac{fista}-\ac{resnn} outperformed the traditional \ac{fista} method in terms of imaging quality and computation time. Our method not only achieved a two-orders-of-magnitude acceleration in inference but also maintained high imaging results with the \ac{ssim} of up to 0.94. Compared with pure neural networks and the standard \ac{fista}, the proposed method achieved the best compromise between computational efficiency and image quality. Moreover, we compared the imaging performance with different datasets: unseen targets dataset, noised raw data with varying \acp{snr}, and raw data performed for different center frequencies. The numerical results verified the robustness and generalization ability of the proposed model and demonstrated the capability of the L-FISTA-ResNet to learn the physical mechanism behind the electromagnetic data.

	\bibliographystyle{IEEEtran}
	\bibliography{IEEEabrv,mybib}

\newpage

 




\vfill

\end{document}